%% file: note1399.tex
\documentclass[twocolumn,showpacs,aps,prl,superscriptaddress,letterpaper]{revtex4}
\usepackage{graphicx}
\usepackage{dcolumn}
\usepackage{amsmath}
\usepackage{epsfig}
\usepackage{multirow}
\usepackage{color}
\usepackage{bm}

\input pubboard/babarsym

\newcommand{\BABARPubYear}    {06}
\newcommand{\BABARPubNumber}  {057}

\newcommand{\SLACPubNumber} {12107}

\newcommand{\dstarstarp}{\ensuremath{D^{**+}}}

\newcommand{\dstarstarz}{\ensuremath{D^{**0}}}
\newcommand{\dstarstar}{\ensuremath{D^{**}}}

\newcommand{\btodpi}{\ensuremath{\Bm\to\Dz\pim}}

\newcommand{\btodspi}{\ensuremath{\Bm \to\Dstarz\pim }}
\newcommand{\btodsspi}{\ensuremath{\Bm \to\dstarstarz\pim}}

\newcommand{\bzbtodpim}{\ensuremath{\Bzb\to\Dp\pim}}
\newcommand{\bzbtodspim}{\ensuremath{\Bzb\to\Dstarp\pim }}
\newcommand{\bzbtodsspim}{\ensuremath{\Bzb\to\dstarstarp\pim}}

\def\D        {\ensuremath{D}\xspace}

\def\Brec  {\ensuremath{\B_{\rm reco}}\xspace}

\def\BbX  {\ensuremath{\Bbar_{ X \pi}}\xspace}

\def\figurebox#1#2#3{%
    \def\arg{#3}%
    \ifx\arg\empty
    {\hfill\vbox{\hsize#2\hrule\hbox to #2{\vrule\hfill\vbox to #1{\hsize#2\vfill}\vrule}\hrule}\hfill}%
    \else
    {\hfill\epsfbox{#3}\hfill}%
    \fi}

\begin{document}

\preprint{\babar-PUB-\BABARPubYear/\BABARPubNumber}
\preprint{SLAC-PUB-\SLACPubNumber}

\begin{flushleft}

%\babar\ Analysis Document \#1399, Version 17 \\
\babar-PUB-\BABARPubYear/\BABARPubNumber\\
SLAC-PUB-\SLACPubNumber\\
%hep-ex/\LANLNumber\\[10mm]
\end{flushleft}

\title
{\large \bf Measurement of the  Absolute Branching Fractions
$\boldmath{B \to D\pi, D^*\pi, D^{**}\pi}$ with a Missing Mass
Method}

\date{\today}% It is always \today, today, but you may specify any date with \date.

\input pubboard/authors_jul2006.tex
\begin{abstract}
We present branching fraction measurements of charged and neutral
\B decays to \D\pim, \Dstar\pim and \dstarstar\pim with a missing
mass method, based on a sample of 231 million \FourS\to\BB pairs
collected by the \babar\ detector at the PEP-II \epem collider.
One of the \B mesons is fully reconstructed and the other one
decays to a reconstructed charged $\pi$ and a companion charmed
meson identified by its recoil mass, inferred by kinematics. Here
$\dstarstar$ refers to the sum of all the non-strange charm meson
states with masses in  the range $2.2-2.8~\gevcc$. We measure the
branching fractions:
\begin{eqnarray}
\BR(\Bm \to \Dz\pi^-)&  =&  (4.49\pm 0.21 \pm 0.23)\times 10^{-3} \nonumber  \\
\BR(\Bm \to \Dstarz\pi^-) & =& (5.13 \pm 0.22 \pm 0.28)\times 10^{-3}  \nonumber  \\
\BR(\Bm \to \dstarstarz\pi^-) & =& (5.50 \pm 0.52 \pm 1.04)\times 10^{-3}\nonumber \\
\BR(\Bzb \to \Dp\pi^-)& =&  (3.03 \pm 0.23 \pm 0.23)\times 10^{-3}\nonumber  \\
\BR(\Bzb \to \Dstarp\pi^-)& =& (2.99 \pm 0.23 \pm 0.24)\times 10^{-3} \nonumber  \\
 \BR(\Bzb \to \dstarstarp\pi^-) &= & (2.34 \pm 0.65 \pm 0.88)\times 10^{-3} \nonumber
\end{eqnarray}
and the ratios:
\begin{eqnarray}
\BR(\Bm\to\Dstarz\pi^-)/\BR(\Bm\to\Dz\pi^-)=1.14 \pm 0.07\pm0.04\  \nonumber\\
\BR(\Bm\to\dstarstarz\pi^-)/\BR(\Bm\to\Dz\pi^-)=1.22\pm0.13\pm0.23\ \nonumber\\
\BR(\Bzb\to\Dstarp\pi^-)/\BR(\Bzb\to\Dp\pi^-)=0.99\pm 0.11\pm0.08\ \nonumber\\
\BR(\Bzb\to\dstarstarp\pi^-)/\BR(\Bzb\to\Dp\pi^-)=0.77\pm0.22\pm0.29\
\nonumber
\end{eqnarray}
The first uncertainty is statistical and the second is systematic.
\end{abstract}

% PACS, the Physics and Astronomy Classification Scheme.
\maketitle

  Our understanding of hadronic \B-meson decays has improved considerably
during the past few years with the development  of the Heavy Quark
Effective Theory (HQET)~\cite{Beneke,Neubert1}
 and the Soft Collinear Effective Theory (SCET)~\cite{Bauer1,Bauer2}.
In these models, and in the  framework of the factorization
hypothesis~\cite{Bauer2,Bauer3}, the amplitude of the $B\to
D^{(*)}\pi$ two-body decay carries information about the
 difference $\delta$ between the
strong-interaction phases of the two isospin amplitudes $A_{1/2}$
and $A_{3/2}$ that contribute~\cite{Rosner,Chiang}. A non-zero
value of $\delta$ provides a measure of the departure from the
heavy-quark limit and the importance of the final-state
interactions in the $D^{(*)}\pi$ system. With the measurements by
the \babar\ \cite{ref:babard0pi0} and
  BELLE \cite{ref:belled0pi0} experiments  of the color-suppressed  $B$ decay
$\Bzb\to\D^{(*)0}\piz$ providing evidence for a sizeable value of
$\delta$, an improved measurement of the color-favored decay
amplitudes ($\Bm\to\D^{(*)0}\pim$  and $\Bzb\to\D^{(*)+}\pim$) is
of renewed interest. In addition, the study of
 $B$~decays into  $D$, $D^{*}$, and $D^{**}$  mesons will
allow tests of  the spin symmetry~\cite{Mannel,NRSX,Mantry,Jugeau}
imbedded in HQET and  of  non-factorizable corrections~\cite{Blok}
that have been assumed to be negligible in the case of the excited
states $D^{**}$ \cite{Neubert2}.

In this paper we present  new measurements of the branching
fractions for the decays \Bm\to\Dz\pim, \Dstarz\pim,
\dstarstarz\pim,~and \Bzb\to\Dp\pim,
 \Dstarp\pim, $\dstarstarp\pim$~\cite{chconj},
based on a missing mass method previously used by
\babar~\cite{worm}. Here \dstarstar\ refers to the sum of all the
non-strange charm meson states with masses in  the range
$2.2-2.8~\gevcc$. This analysis uses \upsbb events in which a \Bp
or a \Bz meson, denoted \Brec, decays into a hadronic final state
and is fully reconstructed. The decays of the recoiling $\Bbar$
into a charged pion and a charmed meson, {\it i.e.} $\Bbar\to \pim
X$, are studied. The charged pion is reconstructed and the mass of
the $X = D, D^*, D^{**}$ is inferred from the kinematics of the
two body \B decay. This method, unlike  the previous exclusive
measurements~\cite{cleo1,babardpi}, does not assume that the
\FourS decays into \Bp and \Bz with equal rates, nor does it rely
on the $D$, $D^*$, or $D^{**}$ decay branching fractions.

The measurements presented here are based on a sample of $231$
million \BB pairs ($210~\invfb$) recorded at the \Y4S resonance
with the \babar\ detector at the \pep2 asymmetric-energy \B
factory at SLAC. The \babar\ detector is described in detail
elsewhere~\cite{det}. Charged-particle trajectories are measured
by a 5-layer double-sided silicon vertex tracker (SVT) and a
40-layer drift chamber (DCH), both operating in a 1.5-T solenoidal
magnetic field. Charged-particle identification is provided by the
average energy loss (\dedx) in the tracking devices and by an
internally reflecting ring-imaging Cherenkov detector. Photons are
detected by a CsI(Tl) electromagnetic calorimeter. Muons are
identified by the  instrumented magnetic-flux return (IFR). We use
Monte Carlo (MC) simulations of the \babar\ detector based on
GEANT4~\cite{GEANT} to optimize selection criteria and determine
selection efficiencies.

\begin{figure}[ht]
\begin{center}
\includegraphics[width=0.80\linewidth]{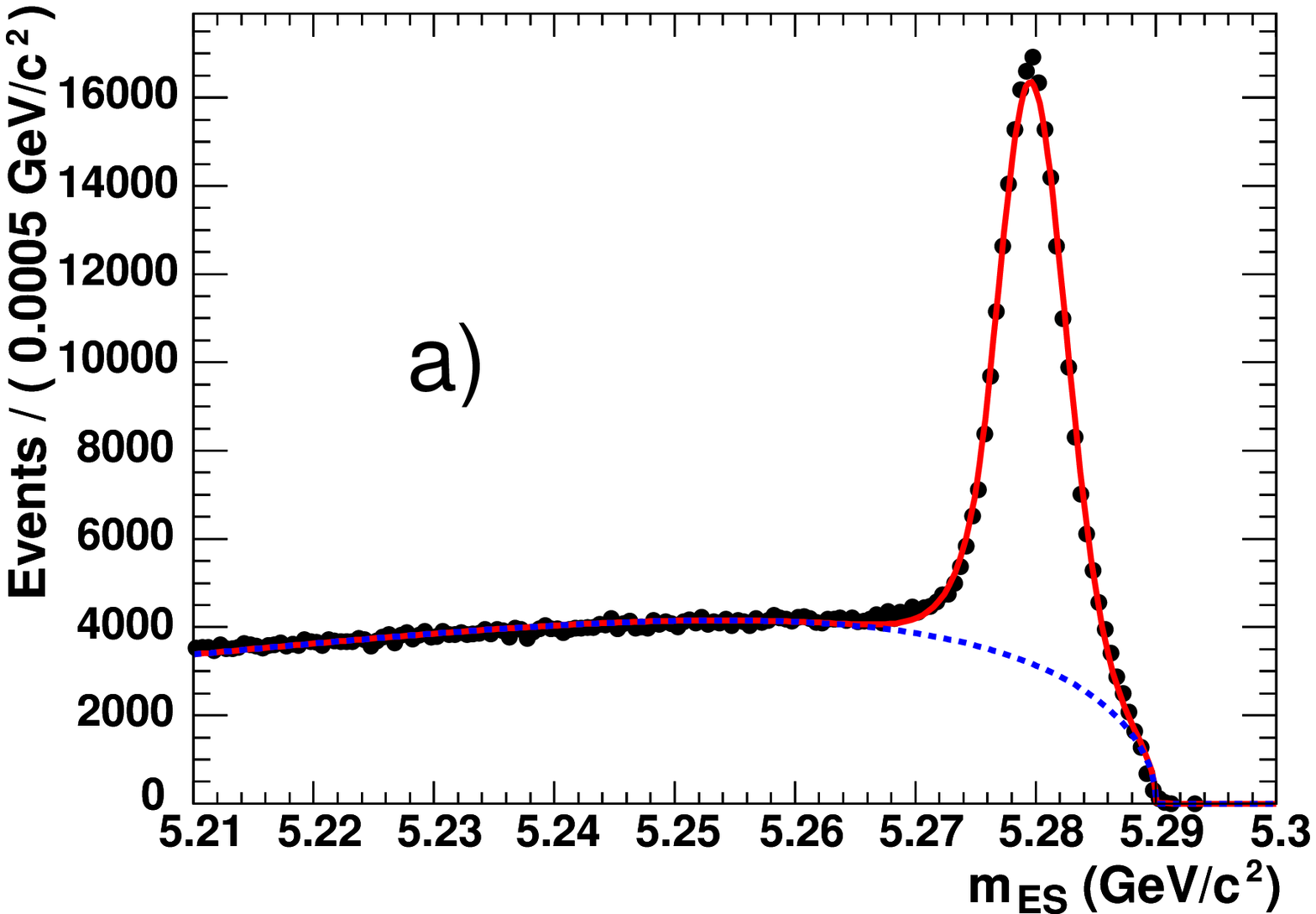}
\includegraphics[width=0.80\linewidth]{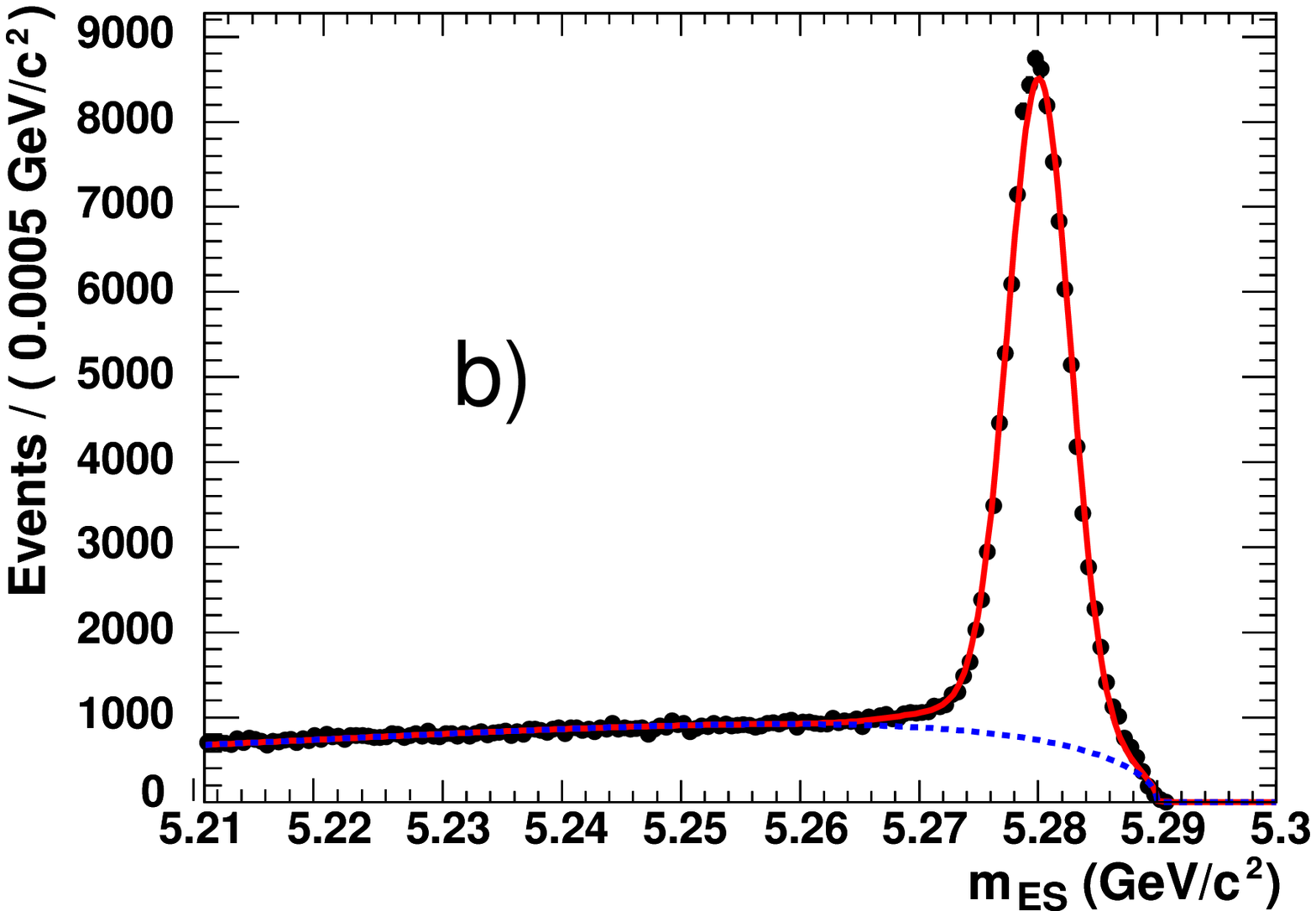}
\caption{\mes spectra of reconstructed (a) \Bp and (b) \Bz
candidates. The solid curve is the sum of the fitted signal and
background whereas the dashed curve is the background component
only.} \label{fig:mes}
\end{center}
\end{figure}

We reconstruct \Bp and \Bz decays (\Brec) in the modes
 $\Bp\to \Db^{(*)0}\pip$, $\Db^{(*)0}\rho^+$, $\Db^{(*)0}a_1^+$,
 and $\Bz\to D^{(*)-}\pip$, $D^{(*)-}\rho^+$, $D^{(*)-}a_1^+$. \Dzb candidates
  are reconstructed in the $\Kp\pim$, $\Kp\pim\piz$,
  $\Kp\pim\pip\pim$,
  and $\KS\pip\pim$ decay channels,
  while \Dm candidates are reconstructed in the $\Kp\pim\pim$ and $\KS\pim$
  modes, and $\KS$ mesons are reconstructed to $\pip\pim$. \Dstar candidates
  are reconstructed in the $D^{*-}\to \Dzb\pim$ and $\Dstarzb\to \Dzb\piz$ decay modes.
  A $3\sigma$ cut is applied on the $\D$  meson mass $m_D$ (and on the
  $\Dstar$-$\D$ mass difference $\Delta m_{\Dstar}$) where $\sigma =
\sigma_{m_D} (\sigma_{\Delta m_{\Dstar}}$) is the resolution on
$m_D$ ($\Delta m_{\Dstar}$)  and is determined from
  data. A vertex fit is performed on $\D$ ($\Dstar$) with the mass
  constrained to the nominal value \cite{PDG2006}.
  Two nearly independent variables are defined to identify the fully reconstructed \B candidates kinematically.
  The first one is the beam-energy substituted mass,
  $\mes =\sqrt{ (s/2 + {\bf p}_{i}\cdot{\bf p}_{B})^{2}/E_{i}^{2}-{\bf p}^{2}_{B}}$,
  where ${\bf p}_{B}$ is the \Brec momentum and $(E_{i},{\bf p}_{i})$ is
  the four-momentum of the initial \epem system, both
   measured in the laboratory frame. The invariant mass of the
   initial \epem system is $\sqrt{s}$.
    The second variable is $\DeltaE = E^*_B-\sqrt{s}/2$,
   where $E^*_B$ is the \Brec candidate energy in the center-of-mass
   frame. To define the \Brec sample (Fig.~\ref{fig:mes}), we require
    $|\DeltaE|<n\,\sigma_{\Delta E}$, where the measured resolutions $\sigma_{\Delta E}$ range from $12$ to $35$ \mev
    and $n=2$ or $3$, both depending on the \Brec mode. The \Brec
    candidate multiplicity is $1.4$ for data as well as for the MC simulation sample.
    For events with more than one candidate, we select the \Brec with the
    best $\chi^2$ defined with the variables $m_D$, $\Delta
    m_{\Dstar}$, and $\DeltaE$. The MC simulation shows that
    the recoil variables are  reconstructed  well within their experimental resolution when using this selection.

The number of \Brec is extracted from  the \mes spectra
(Fig.~\ref{fig:mes}) in the $5.27-5.29~\gevcc$ signal region. The
\mes distribution is fit to the sum of a broad combinatorial
background and  a narrow signal in the  mass interval
$5.21-5.29~\gevcc$. The combinatorial background is described by
an empirical phase-space threshold function~\cite{fargus} and the
signal with a Crystal Ball function~\cite{fcrystalball} which is a
Gaussian function centered at the \B meson
 mass  modified  to  account for
  photon radiation energy-loss. All parameters for the  functions
  describing the \Brec  signal and background distributions are determined from data.
   The measured yields of reconstructed \Bp  and \Bz
  candidates,
      $N_{\Bp}~=~189474~\pm~7487$
   and $N_{\Bz}~=~103169~\pm~3303$, are obtained by subtracting
   the fitted  and the peaking (described below) backgrounds from the total
   number of events found in the signal region.
These $\Brec$ numbers serve as the normalization of all branching
fraction measurements reported in this paper.
The  error is dominated by the systematic uncertainties due to the
fit of the combinatorial background and to the determination of
the peaking background. We assign $2.3\%$ uncertainty to $N_{\Bp}$
and $1.8\%$ to $N_{\Bz}$ as a fit uncertainty, obtained by varying
the lower boundary of the fit interval from $5.20$ to
$5.23~\gevcc$. The contamination of misreconstructed \Bz events in
the \Bp signal (and vice-versa) induces a peaking background near
the \B mass. From the MC simulation, the fraction of \Bz events in
the reconstructed \Bp signal sample is found to be $(3.2 \pm
3.2_{\mathrm {syst.}})\%$ and the fraction of \Bp events in the
reconstructed
 \Bz signal sample $(2.8\pm 2.8_{\mathrm {syst.}})\%$.
 A $100\%$ systematic uncertainty is conservatively assigned to
these numbers taking into account the possible differences in the
reconstruction efficiency in data and MC, as well as the branching
fraction uncertainties for those \B decay modes contributing to
the peaking background. The total systematic
 uncertainties on $N_{\Bp}$ and $N_{\Bz}$ are $3.9\%$ and $3.2\%$, respectively.

In the decay $\FourS \to \Brec \BbX$ where $\BbX $ is the
recoiling $\Bbar$ which decays into $\pim X$, the invariant mass
of the $X$ system is derived from the missing 4-momentum $p_X$
applying  energy-momentum conservation:
\begin{equation}
p_X =p_{\FourS}-p_{\Brec}-p_{\pim}.\nonumber
\end{equation}
The 4-momentum of the \FourS, $p_{\FourS}$, is computed from the
beam energies and $p_{\pim}$ and $p_{\Brec}$ are the measured
4-momenta  of the pion and of the reconstructed $\Brec$,
respectively. The $\Brec$ energy is constrained by the beam
energies. The  $\Bbar\to D \pim$, $\Bbar\to D^* \pim$, or
$\Bbar\to D^{**} \pim$ signal yields peak at the $D$, $D^*$, and
$D^{**}$ masses in the missing mass spectrum, respectively.

The pion candidates, chosen among the tracks that do not belong to
the $\Brec$, are required to have produced  at least 12 DCH hits.
For the charged \Brec, the pion candidate has the opposite sign to
the \Brec. For neutral $\Brec$, because of the $\Bz-\Bzb$ mixing,
the corresponding requirement  is not applied. Muon tracks are
rejected using the IFR information, electrons tracks using the
energy loss in the SVT and the DCH, or the ratio of the
candidate's EMC energy deposition to its momentum ($E/p$). Protons
and kaons are rejected based on informations from the DIRC and
energy loss in the SVT and the DCH. The rejection efficiency is
$97\%$ and there is no peaking trend in the missing mass
distribution from remaining kaons, protons, muons, or electrons.
The pion reconstruction efficiency is determined from the MC
simulation and reported in Table~\ref{tab:branch}.

\begin{figure*}[ht]
\begin{center}
\includegraphics[width=0.450\linewidth]{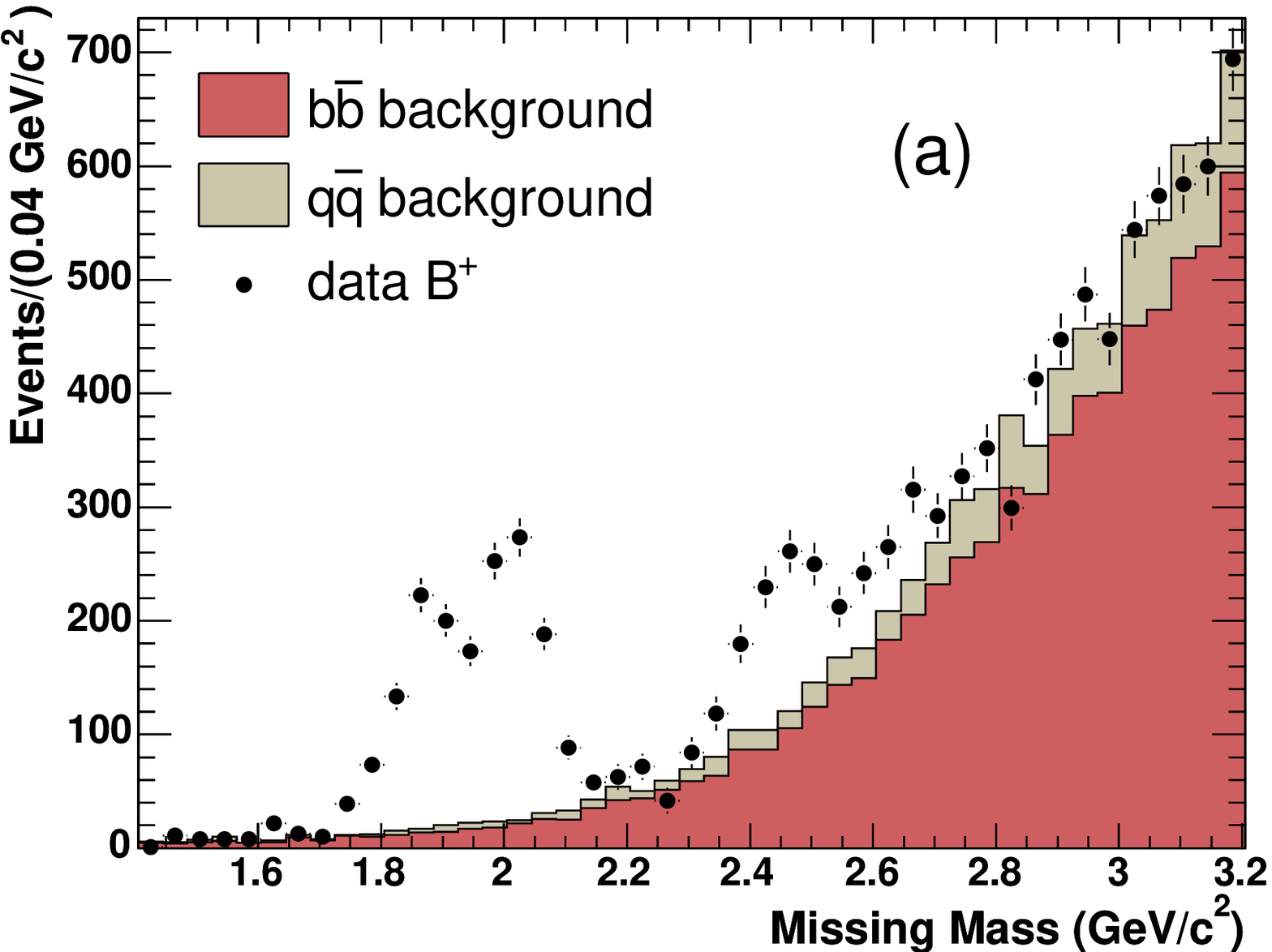}
\includegraphics[width=0.450\linewidth]{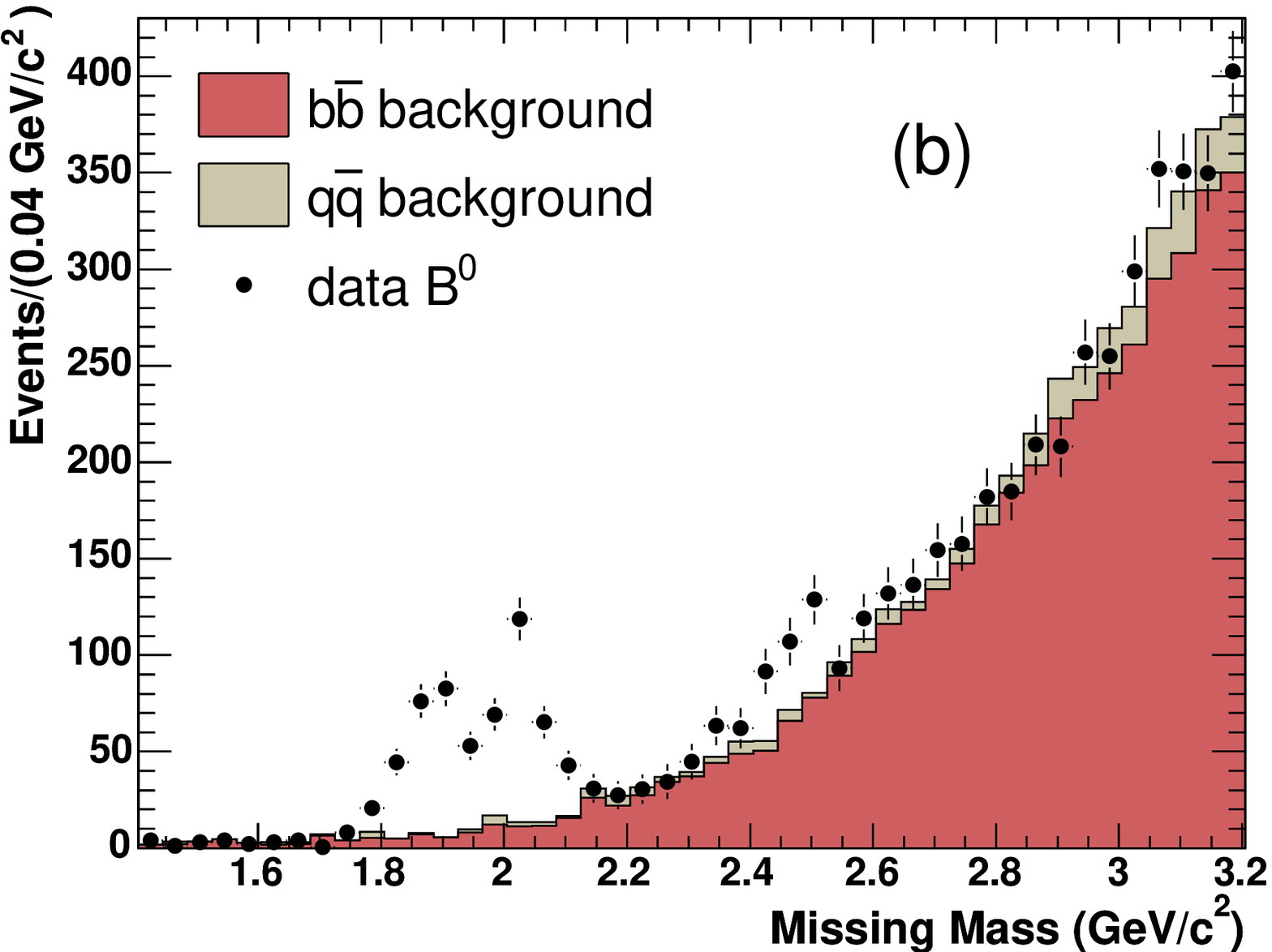}
\includegraphics[width=0.450\linewidth]{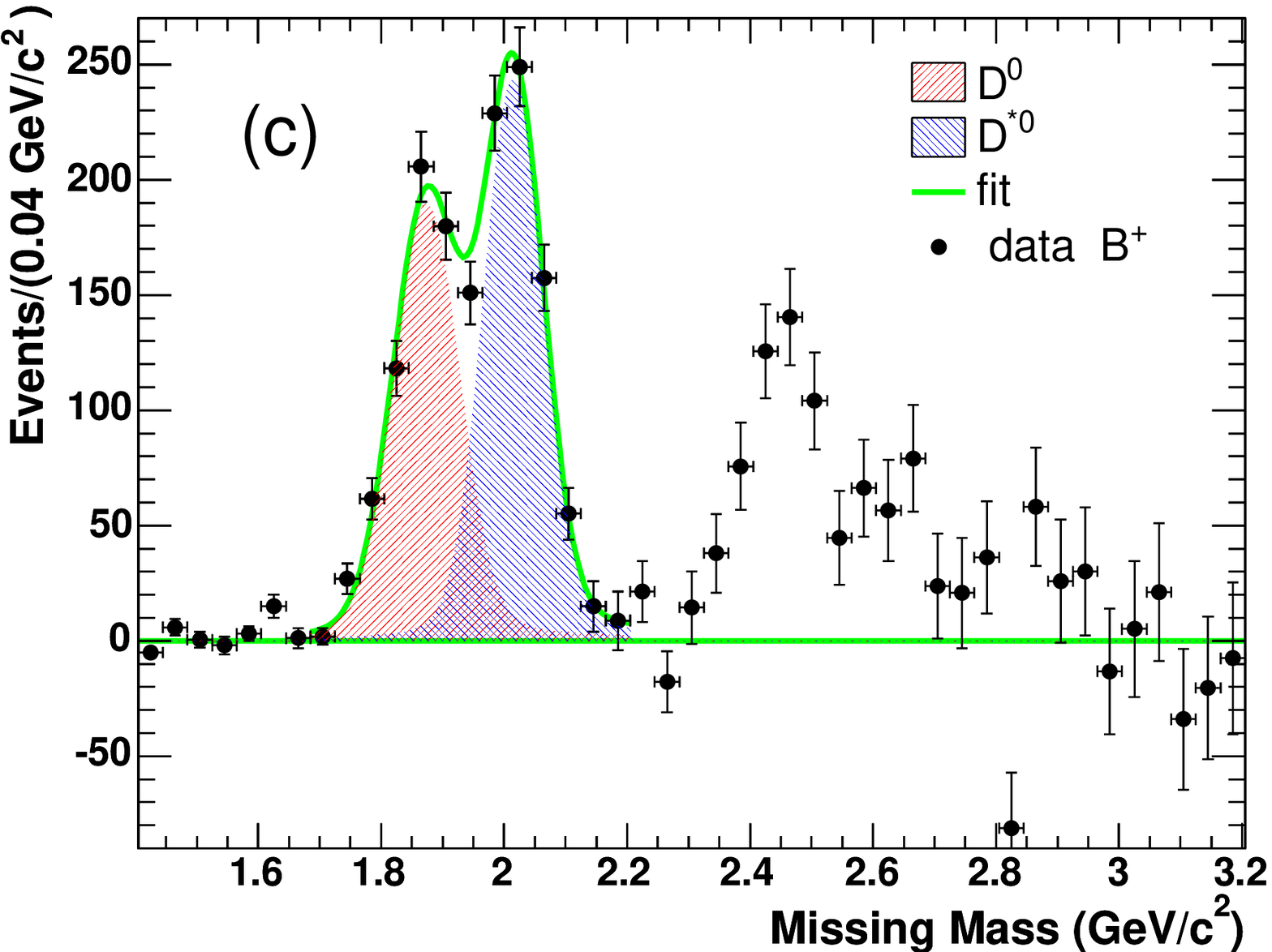}
\includegraphics[width=0.450\linewidth]{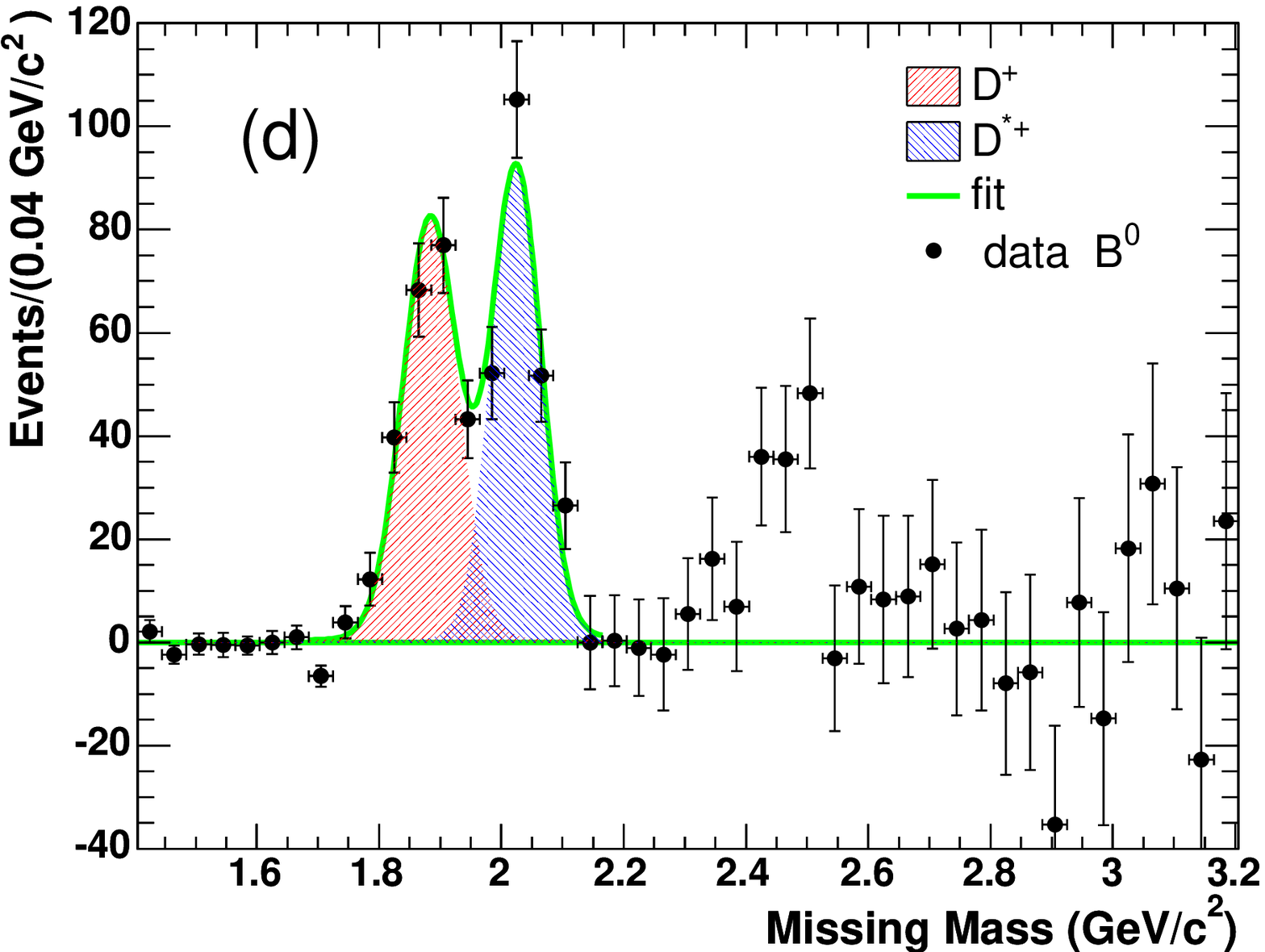}
\end{center}
\caption{ \label{fig:MM} Top: missing mass distributions obtained
in the recoil of $B^+$ (a) and $B^0$ (b). The points with error
bars show the data and the histograms show the
 background contributions (\bbbar and $\qqbar$~$(q=c,u,d,s)$) predicted by the MC simulation.
Bottom: background-subtracted missing mass spectra for $B^+$ (c)
and $B^0$ (d). The curves show the result of the fits to the
$D\pi$ and $D^*\pi$ components.}
\end{figure*}

The signal yields for the different decay modes are extracted from
the missing mass spectra. The data distributions and the $\bbbar$
and the $\qqbar$~$(q=c,u,d,s)$ background expectations are shown
in Figs.~\ref{fig:MM}(a) and~\ref{fig:MM}(b). The shape of the
background is taken from MC and the normalization is scaled to
match the data in the sideband region $2.8-3.2~\gevcc$. The error
on the background normalization is  $2\%$. This is determined
using the statistical errors of MC and data samples. The
background subtracted missing mass distributions are shown in
Figs.~\ref{fig:MM}(c) and~\ref{fig:MM}(d).

The $\D\pi$ and $\Dstar\pi$ signal yields are extracted  by a
$\chi^2$ fit to the background subtracted missing mass
distribution in the range $1.65-2.20~\gevcc$. The $\D\pi$ and
$\Dstar\pi$ components are each modeled by a sum of two Gaussian
functions, to account for tails in the mass distributions. The
 parameters are  $m_i^{\D^{(*)}}$ and
$\sigma_i^{\D^{(*)}}$  for the $\D$  and $\Dstar$ resonances,
where the index $i=1,2$ corresponds to the first and second
Gaussian. In the fit, the central values $m_i^{\D} $ and the
$\sigma_i^{\D}$ are  free parameters, while for the $\Dstar$
 the variances are constrained by the ratios
 $\sigma_i^{\Dstar}$$/$$\sigma_i^{\D}=0.900\pm0.015$, as determined from MC
 simulation, while  the central values differences $m_i^{\Dstar}-m_i^{\D}$ are fixed to
 $0.1421~\gevcc$ and to $0.1406~\gevcc$  for $\Bp$ and
 $\Bz$, respectively, corresponding to the world average $\D$ and
 $\Dstar$ mass differences~\cite{PDG2006}.
\par The $D^{**}$ yields are defined as the excess of candidates in
the missing mass range $2.2-2.8~\gevcc$, and the $\Bbar\to
D^{**}\pim$ branching fractions refer to the contributions of all
non-strange charm meson states in the same region. The range is
chosen in order to maximize the acceptance to the four P-wave
\dstarstar\ states predicted by the theory given the $34$ \mevcc
mass resolution, determined from MC simulation, in the same
region. The well-known narrow $D_1$ and $D_2^*$
states~\cite{PDG2006} are fully contained in this range, and more
than $90\%$ of the broad $D_0$ and $D'_1$, are covered if measured
masses and widths ~\cite{belledstdst,FOCUS:FermiLab} are used. The
event yields, the efficiencies, and the resulting branching
fractions are reported in Table~\ref{tab:branch}.

\begin{table}[htb]
\begin{center}
\caption{{ \label{tab:branch} Signal yields, efficiencies and
branching fractions for $\Bbar\to D\pim$, $\Bbar\to \Dstar\pim$
and $\Bbar\to \dstarstar\pim$. The first error is statistical
except for the efficiencies for which it is mainly systematic. The
 second error on the branching fractions
is systematic. The $\Bbar\to \dstarstar\pim$ branching fractions
are given for the $2.2-2.8~\gevcc$ mass range which in addition to
the P-wave states may include some yet unknown charm meson
states.}}
\begin{tabular}{lccc}
  \hline \hline

  Decay mode&~~~~Yield~~~~ &Efficiency&\BR ($10^{-3}$)\\

\hline
  \btodpi    & 677 $\pm$ 32   &                   & 4.49$\pm$0.21$\pm$0.23 \\
  \btodspi   & 774 $\pm$ 33   & 0.796$\pm$0.007~~~& 5.13$\pm$0.22$\pm$0.28 \\
  \btodsspi  & 829 $\pm$ 78   &                   & 5.50$\pm$0.52$\pm$1.04 \\

\hline
  \bzbtodpim  & 248 $\pm$ 19  &                    & 3.03$\pm$0.23$\pm$0.23 \\
  \bzbtodspim & 245 $\pm$ 19  &  0.793$\pm$0.007~~~& 2.99$\pm$0.23$\pm$0.24 \\
  \bzbtodsspim & 192 $\pm$ 54 &                    & 2.34$\pm$0.65$\pm$0.88 \\
\hline \hline
\end{tabular}
\end{center}
\end{table}

 The uncertainty related to $\pi$
reconstruction efficiency is due to the MC sample statistics and
the systematic uncertainty on  track reconstruction and particle
identification algorithms. The uncertainty due to the yield
extraction is estimated by fitting the MC sample. The difference
between the MC and the data fitted yields is found to be
consistent with zero and the statistical errors are
 taken as a systematic error. We evaluate the uncertainty
 on the missing mass resolution in the $D\pi$ and $\Dstar\pi$ yield
 extraction  by varying by one standard deviation the ratio
 $\sigma_i^{\Dstar}$$/$$\sigma_i^{\D}$ while $\sigma_2^{\Dstar\ }$ and $m_2^{\Dstar\ }$
 are let free.
 The  difference in the yield is taken as systematic uncertainty.
 The uncertainty related to the subtraction of the background  is
determined by varying the branching fraction of the different
background components within the uncertainties of the most recent
measurements~\cite{PDG2006} and taking into account the error on
the background normalization. Due to the threshold shape of some
of the background components and the fast varying combinatorial
background, $\B \to \D^{**}\pi$ branching fractions have larger
systematic errors than $\B \to \D\pi$ and $\B \to \D^{*}\pi$
branching fractions. The summary of these systematic uncertainties
 is reported in Table~\ref{tab:BchTotSys}.\\

\begin{table*}[htb]
\begin{center}
\caption{{ \label{tab:BchTotSys} Total relative systematic
uncertainties for the  branching fractions
\BR(\Bm~\to~(\Dz,~\Dstarz,~\dstarstarz)\pim) and
\BR(\Bzb~\to~(\Dp,~\Dstarp,~\dstarstarp)\pim). }}
\begin{tabular}{lcccccc}
  \hline \hline
  Syst. Source &\btodpi  &  \btodspi  & \btodsspi &\bzbtodpim   &  \bzbtodspim & \bzbtodsspim\\
\hline
  $N_B $          &3.9\%       &3.9\% & 3.9\%   &3.2\%       &3.2\%& 3.2\% \\
  Efficiency      &0.9\%       &0.9\% & 0.9\%   &0.9\%       &0.9\%& 0.9\% \\
  Yield extraction           &2.7\%       &2.7\% & 5.1\%   &5.4\%       &5.1\%& 5.9\% \\
  Missing mass resolution       &0.9\%       &0.8\% & -   &1.9\%       &1.1\%& - \\
  Background subtraction     &1.6\%       &2.3\% & 17.7\%  &3.7\%       &5.4\%& 37.1\% \\
\hline
Total             &5.2\%       &5.4\% & 18.9\%  &7.6\%       &8.2\%& 37.7\% \\
\hline \hline
\end{tabular}
\end{center}
\end{table*}

Using the measured branching fractions we compute the following
ratios:
\begin{eqnarray}
\BR(\Bm\to\Dstarz\pi^-)/\BR(\Bm\to\Dz\pi^-)=1.14 \pm 0.07\pm0.04,\  \nonumber\\
\BR(\Bm\to\dstarstarz\pi^-)/\BR(\Bm\to\Dz\pi^-)=1.22\pm0.13\pm0.23,\ \nonumber\\
\BR(\Bzb\to\Dstarp\pi^-)/\BR(\Bzb\to\Dp\pi^-)=0.99\pm 0.11\pm0.08,\ \nonumber\\
\BR(\Bzb\to\dstarstarp\pi^-)/\BR(\Bzb\to\Dp\pi^-)=0.77\pm0.22\pm0.29.\
\nonumber
\end{eqnarray}
The first uncertainty is statistical and the second is systematic.
In addition to the cancellation of many of the systematic errors,
the  ratios are insensitive to the absolute normalization scale.

 In summary, we  have measured the branching fractions for the decays
$\btodpi$, $\btodspi$, $\btodsspi$, $\bzbtodpim$, $\bzbtodspim$,
and $\bzbtodsspim$, using a missing mass method. This measurement
does not assume that the \FourS decays into \Bp and \Bz with equal
rates, nor does it rely on the $D$, $D^*$, or $D^{**}$
intermediate branching fractions. The results
 for $\BR(\B\to\D\pi^-)$ and $\BR(\B\to\Dstar\pi^-)$ are compatible
 with previous world averages~\cite{PDG2006}.  We have  extracted a new result
 for $\BR(\B\to\dstarstar\pi^-)$ branching fractions where
 $\dstarstar$ excited  states correspond to the yield measured in the mass range
 $2.2-2.8~\gevcc$. The isospin study \cite{Rosner,Chiang} will become competitive with the exclusive
  measurements \cite{babardpi} if the statistical error is reduced by a factor of
  $2$.  With regard to spin symmetry, the values measured for the ratios $\BR(\Bm\to\Dstarz\pi^-)/\BR(\Bm\to\Dz\pi^-)$ and
  $\BR(\Bzb\to\Dstarp\pi^-)/\BR(\Bzb\to\Dp\pi^-)$ are close to 1, as predicted by different theoretical
models~\cite{Mannel, NRSX, Mantry, Jugeau, Blok}, and their
precision is comparable or better than the current world
averages~\cite{PDG2006}.

\input pubboard/acknowledgements

\end{document}

%% file: pubboard/authors_jul2006.tex
%% author list as of 01-Jul-2006 (596 authors)
%
\author{B.~Aubert}
\author{M.~Bona}
\author{D.~Boutigny}
\author{F.~Couderc}
\author{Y.~Karyotakis}
\author{J.~P.~Lees}
\author{V.~Poireau}
\author{V.~Tisserand}
\author{A.~Zghiche}
\affiliation{Laboratoire de Physique des Particules, IN2P3/CNRS et Universit\'e de Savoie,
 F-74941 Annecy-Le-Vieux, France }
\author{E.~Grauges}
\affiliation{Universitat de Barcelona, Facultat de Fisica, Departament ECM, E-08028 Barcelona, Spain }
% ...
\collaboration{The \babar\ Collaboration}
\noaffiliation

%% file: pubboard/acknowledgements.tex
We are grateful for the 
extraordinary contributions of our \pep2\ colleagues in
achieving the excellent luminosity and machine conditions
that have made this work possible.
The success of this project also relies critically on the 
expertise and dedication of the computing organizations that 
support \babar.
The collaborating institutions wish to thank 
SLAC for its support and the kind hospitality extended to them. 
This work is supported by the
US Department of Energy
and National Science Foundation, the
Natural Sciences and Engineering Research Council (Canada),
Institute of High Energy Physics (China), the
Commissariat \`a l'Energie Atomique and
Institut National de Physique Nucl\'eaire et de Physique des Particules
(France), the
Bundesministerium f\"ur Bildung und Forschung and
Deutsche Forschungsgemeinschaft
(Germany), the
Istituto Nazionale di Fisica Nucleare (Italy),
the Foundation for Fundamental Research on Matter (The Netherlands),
the Research Council of Norway, the
Ministry of Science and Technology of the Russian Federation, 
Ministerio de Educaci\'on y Ciencia (Spain), and the
Particle Physics and Astronomy Research Council (United Kingdom). 
Individuals have received support from 
the Marie-Curie IEF program (European Union) and
the A. P. Sloan Foundation.

%% file: note1399.bbl
\begin{thebibliography}{99}
\bibitem{chconj} Charge conjugate relations are assumed throughout
this paper.
\end{thebibliography}
